\documentclass[aps,prl,a4paper,floatfix,showkeys,reprint,showpacs]{revtex4-1}
\usepackage[prependcaption,textsize=tiny]{todonotes}
\usepackage{amsmath}%
\usepackage{amsfonts}%
\usepackage{amssymb}%
\usepackage{graphicx}
\usepackage{hyphenat}
\usepackage{siunitx}
\usepackage{color}
\usepackage[normalem]{ulem}
\usepackage{ifthen}
\usepackage{soul}

\sethlcolor{orange}

\newcommand{\fig}[1]{Fig. {\ref{#1}}}

\newcommand{\unit}[1]{\,\,{\si{#1}}}

\newboolean{showcomment}
\setboolean{showcomment}{true}
\ifshowcomment %
	\newcommand{\pico}[1]{\textcolor{red}{PICO:\textbf{#1}}}
	
	\newcommand{\delete}[1]{\textcolor{red}{\sout{#1}}}
	
\else
  \newcommand{\pico}[1]{}
	
	\newcommand{\delete}[1]{}
	
	\renewcommand{\todo}[1][2]{}
\fi

\begin{document}
\renewcommand{\figurename}{Figure}

\title{Ultrafast coherent dynamics of a photonic crystal all-optical switch} 



\author{Pierre Colman}
\affiliation{DTU Fotonik, Technical University of Denmark, Kongens Lyngby, Denmark}
\affiliation{IEF Institut d'Electronique Fondamentale, Universit\'{e} Paris-Sud, Orsay, France}

\author{Per Lunnemann}
\affiliation{DTU Fotonik, Technical University of Denmark, Kongens Lyngby, Denmark}

\author{Yi Yu}
\affiliation{DTU Fotonik, Technical University of Denmark, Kongens Lyngby, Denmark}

\author{Jesper M{\o}rk}
\email[]{jesm@fotonik.dtu.dk}
\affiliation{DTU Fotonik, Technical University of Denmark, Kongens Lyngby, Denmark}




\date{\today}

\begin{abstract}
We present pump-probe measurements of an all-optical photonic crystal switch based on a nanocavity, resolving fast coherent temporal dynamics. The measurements demonstrate the importance of coherent effects typically neglected when considering nanocavity dynamics. In particular, we report the observation of an idler pulse. The measurements are in good agreement with a theoretical model that allows us to ascribe the observation to oscillations of the free carrier population in the nanocavity. The effect opens perspectives for the realization of new all-optical photonic crystal switches with unprecedented switching contrast. 
\end{abstract}

\maketitle 


Over the last decade there has been significant progress in integrated optics in terms of decreasing both footprint and energy consumption. Photonic solutions are thus increasingly becoming a credible alternative to electrical signal processing. The control of light with highly efficient all-optical functions, such as active switching/gating operations or the use of integrated add/drop channels, is of utmost importance. In order to cope with the constraints related to the dense integration of numerous all-optical functions on a single integrated photonic chip (IPC), the total energy consumption devoted to each individual function must be of the order of a few fJ/bit \cite{Miller09}. Planar photonic crystal (PhC) cavities are promising candidates for the realization of all-optical switching operations thanks to their small volume, high quality factor, and compatibility with complementary metal-oxide-semiconductor (CMOS) technology \cite{tanabe05, Shinya2008,husko2009,Yu:13,6671404,Notomi:10b}. In particular, the report of 10~dB switching contrast with a record low operating energy of $2.88\unit{fJ/bit}$ is noteworthy \cite{Notomi:10b}.

\begin{figure}
\centering
\includegraphics[width=1\columnwidth]{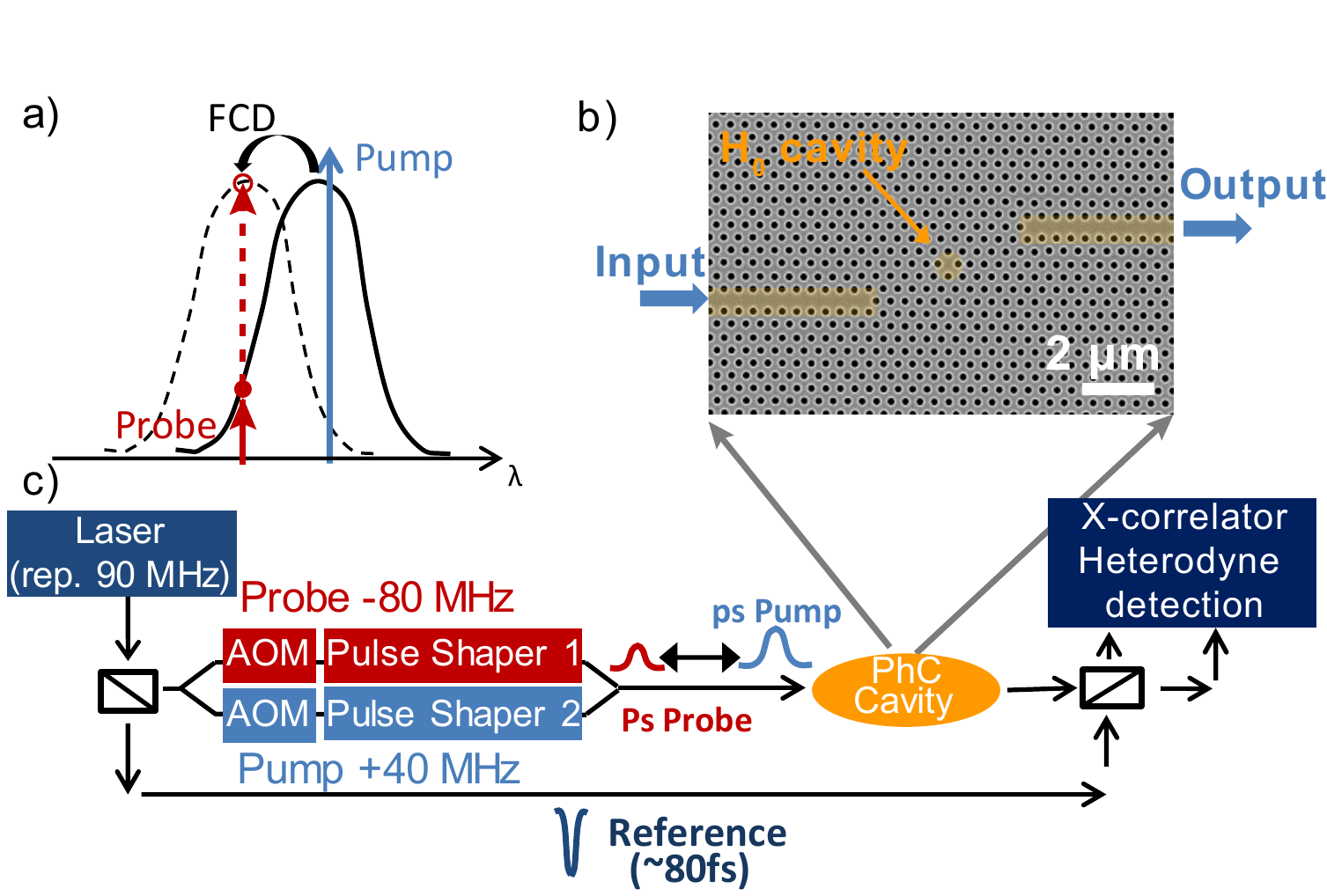}
\caption{a) Principle of operation of a PhC cavity all-optical switch. b) SEM image of the $H_0$ cavity and its access waveguides. c) Schematic of the experiment. The picosecond pump and probe are carved from the same 80~fs reference pulse. Heterodyne detection using short reference pulses allows characterizing pulses with energy as low as 40~aJ/pulse and achieving 150~fs deconvolution-free temporal resolution.}
\label{Fig_I}
\end{figure}

Classical schemes for all-optical switching using a PhC cavity involve the dynamical control of the cavity resonance via a pump pulse \cite{tanabe_optexp05,tanabe05,notomi2005}, which shifts the cavity's resonance, thus controlling the transmission of a subsequent probe (see \fig{Fig_I}b)). The cavity resonance can be changed either by the Kerr effect \cite{Lan04}, or through the dispersion caused by free carriers (FCD) \cite{fushman2007} generated by the absorption of the pump. The latter process is usually preferred as it has the advantage of building up over time and therefore requires less pump power. 
Thus far, the lowest switching energy was obtained by taking advantage of a combination of linear absorption and nonlinear two-photon absorption (TPA) \cite{Notomi:10b, Notomi:10} in a configuration that benefits from the band filling dispersion. The band filling dispersion adds up to the free carriers dispersion (FCD) to give rise to a stronger resonance shift \cite{Bennet90}. However these demonstrations require complex material engineering, and showed limitations in terms of switching speed due to a long free carrier lifetime. In particular, a long carrier lifetime gives rise to strong patterning effects when operated at a high rate \cite{Yu:13,Moille2016}.
The use of a small cavity with high-Q factor favors the use of TPA-based carrier generation \cite{Bravo-Abad2007} as does the use of shorter pulses.
Although TPA-based switching usually requires higher switching energy than switching based on linear absorption, the energy consumption may be further reduced using waveguide-cavity designs exhibiting Fano resonances \cite{sharp02,HuskoAPL07,Yi:OL15}, where a nonlinear regeneration mechanism has been shown to improve the response \cite{Yi_LPR15}.

Despite a good understanding of the dynamics of PhC switches on the long time scale \cite{derossi_PRA2009,weidner07,Yu:13}, the switching dynamics on short timescales as well as the possible role of coherent effects has so far received little attention. In addition to the resonance shift, caused by the FCD, the PhC cavity is also known to induce, during the first stage of its nonlinear evolution, an adiabatic frequency shift of the light \cite{tanabe06,preble07,mosk:10} and to undergo rapid changes of its refractive index \cite{Kuipers}. The impact of these effects on the switching dynamics has not been considered. In particular, the coherent interaction of the pump and probe pulses has to the best of our knowledge been neglected so far in all models and interpretations of TPA based switching dynamics. Nevertheless, it is well-known that coherent effects play an important role on the dynamics of active semiconductor waveguides at short time-scales \cite{Mecozzi:96}. Photonic semiconductor structures exhibiting nonlinear absorption should not be exempt from such effects. A good understanding of the different mechanisms governing the ultrafast switching of a semiconductor micro cavity is indeed essential for the optimization of future all-optical switches. 

In this work we utilize a pump-probe technique based on a heterodyne detection scheme \cite{Hall1992} that allows measuring the temporal trace of atto-joule probe pulses
spectrally overlapping with a pump pulse, while having  the unprecedented temporal and energy resolution of $\leq100\unit{fs}$ and $\leq40\unit{aJ/pulse}$. We investigate the switching dynamics in the ultra fast regime and show in particular that the fast modulation of the cavity gives rise to large parametric gain ($>
10\unit{dB}$) induced by population oscillations of the free carrier plasma. This coherent process adds to the classical switching mechanism that is observed for long time-scales. By generalizing the classical temporally coupled mode theory (TCMT), used so far to describe switching dynamics \cite{derossi_PRA2009,Yu2014a,heuck_APL13}, to include coherent pump-probe interactions, very good agreement between experiment and theory is obtained.



We consider a photonic crystal InP membrane with an $\textrm{H}_0$ nanocavity \cite{nozaki2006} coupled with access waveguides as seen in \fig{Fig_I}-b. InP exhibits TPA at the excitation wavelength of $1.55\unit{\micro\meter}$ (gap is 1.34~eV \cite{HandbookChemistry}). The quality factor of the cavities that we studied is in the range $Q=1000 - 3000$, corresponding to a cavity lifetime of the same order as the fast diffusion time (about 3~ps, \cite{Yu:13}) of the free carriers. Note that cavities with ultra-high quality factor, working at lower speed, can experience also other phenomena like for example regenerative oscillations \cite{Cazier:13,CWW_APL14,CWW_Naptphot12} or self-pulsing \cite{biancalana:11}, but they happen on a much longer time scale than the effects investigated here.

In previous experiments \cite{Yu:13,Yu2014a}, only the total (integrated) probe output energy is detected. Here, we retrieved the temporal shape of the transmitted probe. Using pulse shapers, we shaped Fourier limited gaussian picosecond pump and probe pulses matching the PhC cavity spectral width while keeping the reference as short as possible (0.1~ps).  With the short  duration of the reference pulse, scanning the reference-probe delay stage is nearly equivalent to recording the probe envelope without convolution issues that usually smear the temporal features in cross-correlation measurements \cite{heuck_APL13}. As we shall see, the temporal resolution becomes important for observing the reported features of this paper.

\begin{figure}
\centering
\includegraphics[width=8.5cm]{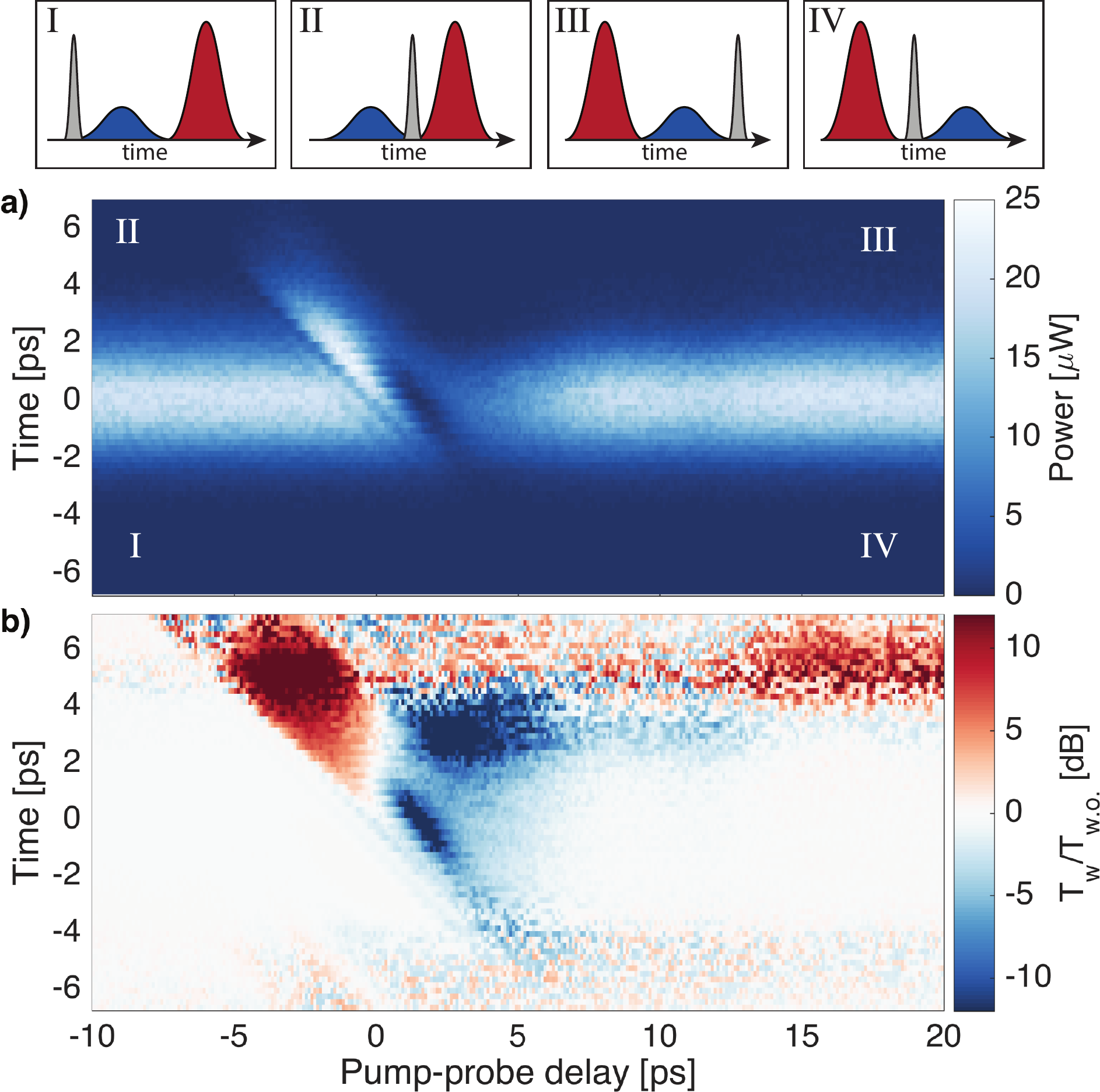}
\caption{Switch-off dynamics. a)  Measured probe trace as a function of pump-probe delay. Y-axis ('Time') corresponds to the reference-probe delay: the ultra-short duration of the reference pulse implies near-instantaneous gating.  The temporal sequences of the reference (grey), pump (red) and probe (blue) pulses at points I-IV are illustrated. b) Measured differential probe transmission in dB (transmission with pump on divided by transmission with pump off). Pump and probe pulses are set on resonance with the cold cavity and have temporal widths of about 1ps and 2ps respectively.}
\label{Fig_III} 
\end{figure}

We consider the PhC cavity switch in a switch-off configuration, where probe and pump are set on resonance with the unperturbed "cold" cavity. As the cavity shifts under the action of the pump, the probe transmission drops. For each pump-probe delay, we retrieved the transmitted probe envelope simply by scanning the reference stage. Differential transmission is obtained by recording the probe trace both with and without the pump present. The resulting transmission map is presented in \fig{Fig_III}a).
For large negative pump-probe delays, the probe arrives well before the pump. Its shape is nearly symmetric, indicating a good match between the probe duration and the cavity response. Moreover, its transmission is not perturbed by the pump. For positive pump-probe delays, the probe is first suppressed and subsequently recover within $\sim10$~ps. This is associated with a blue shift of the cavity resonance caused by the generation of free carriers, leading to a reduction of the probe transmission and followed by relaxation back towards equilibrium \cite{heuck_APL13,Yu:13}. 
However, a very surprising effect appears for negative pump-probe delays, where the pump pulse excites the cavity just after the probe has entered: it induces a very strong {\em increase} in the transmission of the probe signal resulting in the appearance of a secondary peak. This effect has to our knowledge not been observed before. Previous models developed for long pulses and neglecting coherent mixing effects only account for a reduction in transmission. 
The strength of the effect is particularly well illustrated in \fig{Fig_III}b) showing the transmission ratio of the probe with and without the pump. Here, blue (red) areas indicate transmission suppression (enhancement). For pump delays of $\sim-3$~ps, the transient transmission is increased by more than $10\unit{dB}$. 

We ascribe the transient increase of the probe transmission to parametric gain caused by four wave mixing that amplifies the tail of the probe pulse as soon as the pump starts entering the cavity. This interpretation is tested by investigating the presence of an idler signal at the same moment the transient increase of the probe is observed. The heterodyne detection technique used in our experiments is practical in this aspect since the idler signal can be selected by filtering at the proper RF beating frequency \cite{Mecozzi:96}.
\begin{figure}
\centering
\includegraphics[width=8.5cm]{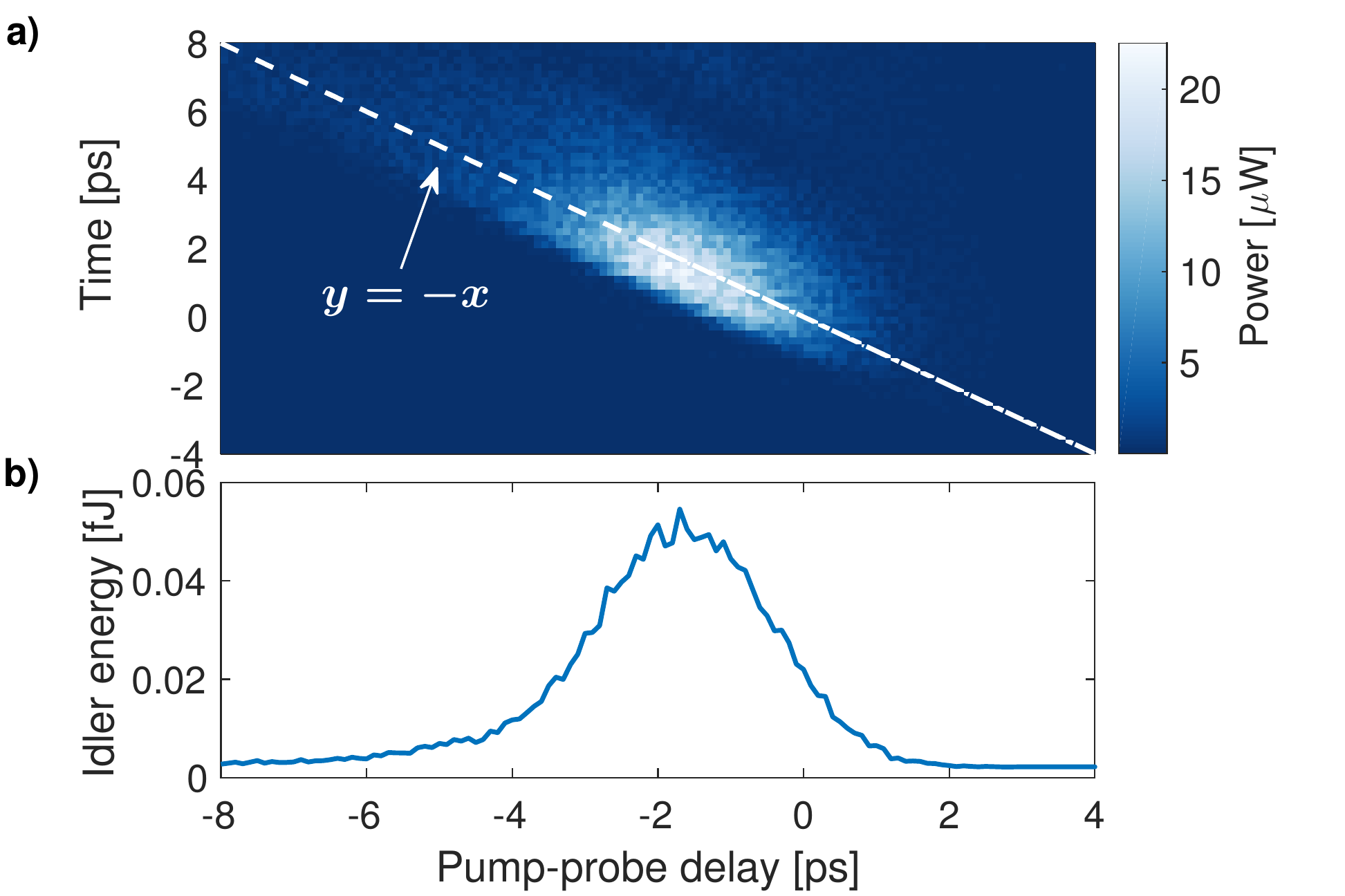}
\caption{a) Measured envelope of idler pulse, obtained by monitoring the signal at the beating frequency of 21.54~MHz, as function of pump-probe delay and reference delay. Dashed line indicates coincidence of the pump and reference pulse peaks. b) Total idler energy versus pump-probe delay.}
\label{Fig_IV} 
\end{figure}
Indeed the signal at the idler frequency, presented in \fig{Fig_IV},  has a similar shape as the secondary peak in \fig{Fig_III}-a), and furthermore coincides with the pump arrival time (thick dashed white line). The total idler energy is plotted in \fig{Fig_III}-b). 

\begin{figure}
\centering
\includegraphics[width=.9\columnwidth]{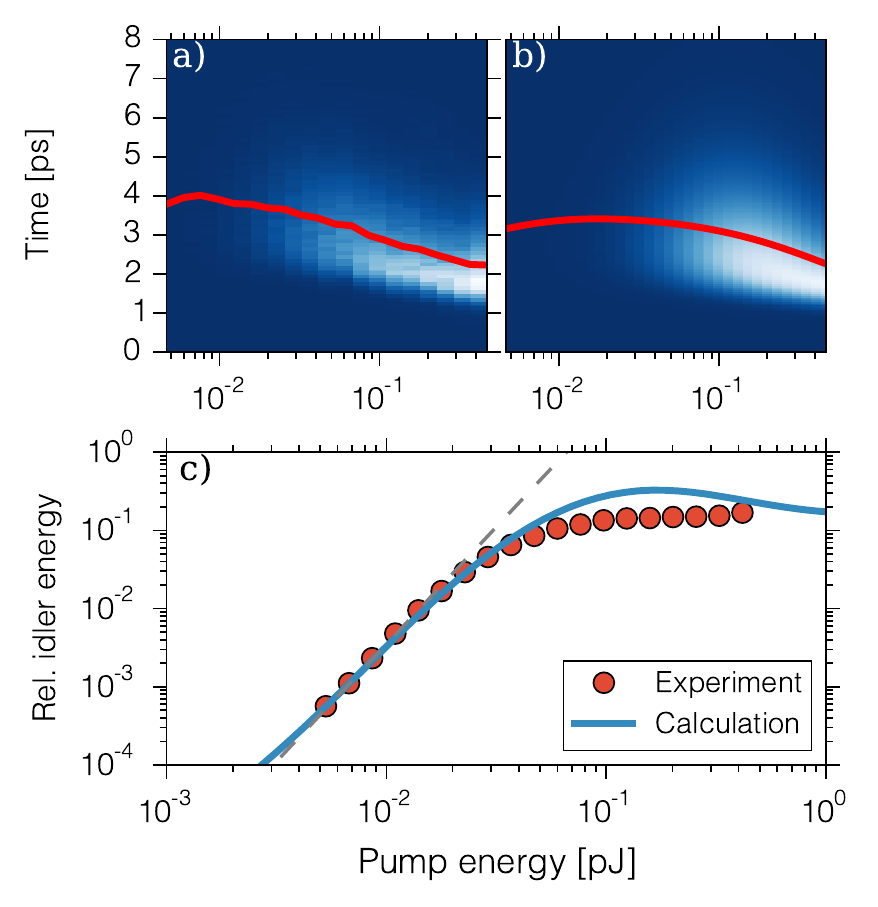}
\caption{a) Measured envelope of idler pulse as a function of coupled pump pulse energy and reference delay ('Time'). The pump-probe delay was set at -2~ps. Red line indicates the temporal mean (first moment). b) Corresponding theoretical calculation. c) Idler energy relative to the input probe pulse energy, obtained by integration of results depicted in a) and b) along the time axis. Red circle markers indicate experimental data and solid blue line is calculated results. Dashed line is a guide to the eye indicating a third order power scaling ($y=ax^3$).\label{Fig_V} }
\end{figure}

We also recorded in \fig{Fig_V}-a) the idler's temporal envelope for a fixed pump-probe delay of -2~ps and varying pump pulse energies. We also show theoretical results, \fig{Fig_V}-b), obtained by generalizing coupled-mode theory \cite{Yu:13,heuck_APL13,derossi_PRA2009} to include coherent wave mixing. 
As the pump energy is increased, the envelope gradually shifts towards earlier times. The corresponding total idler energy relative to the input probe energy is shown in \fig{Fig_V}-c). Note that the experimental measurements are scaled to correspond with the theoretical values for the lowest energy. The energy of the idler pulse quickly increases with pump energy and saturates for pump energies above 100-150~fJ. For the lowest pump powers, the idler energy clearly scales with the pump power to the power of 3, as indicated with the dashed line. Such a scaling differs from the square scaling that is expected for Kerr ($\chi^{(3)}$)-based parametric amplification. From \fig{Fig_V}a) we oberserve, for the largest pump energy, that the maximal idler signal is not obtained at a time identical to the arrival time of the pump (2 ps), rather the maximal idler signal is obtained at slightly earlier times (1.8~ps). As the pump enters the cavity, the induced resonance shift affects both the phase matching between the pump, probe and idler as well as the probe's in and outcoupling of the cavity. E.g. for moderate pump energy, all three waves remain on resonance within the cavity, the parametric process builds up and results in a large output idler energy. For a large pump energy the shift of the resonance is too large and the idler/probe leaves the cavity before the pump has fully entered. Therefore, the parametric gain saturates when the pump pulse reaches an energy of  100-150fJ. The higher intrinsic gain due to higher pump power is compensated by a decreased interaction time. This also results in a decrease with pump power of the arrival time of the idler pulse.
Quantitative differences between theory and experiment appear for the highest pump energies and are attributed to the choice of model parameters in a large parameter space, as well as the limitations of the simplified model where thermal effects are not included.

\begin{figure}\textbf{}
\centerline{\includegraphics[width=7cm]{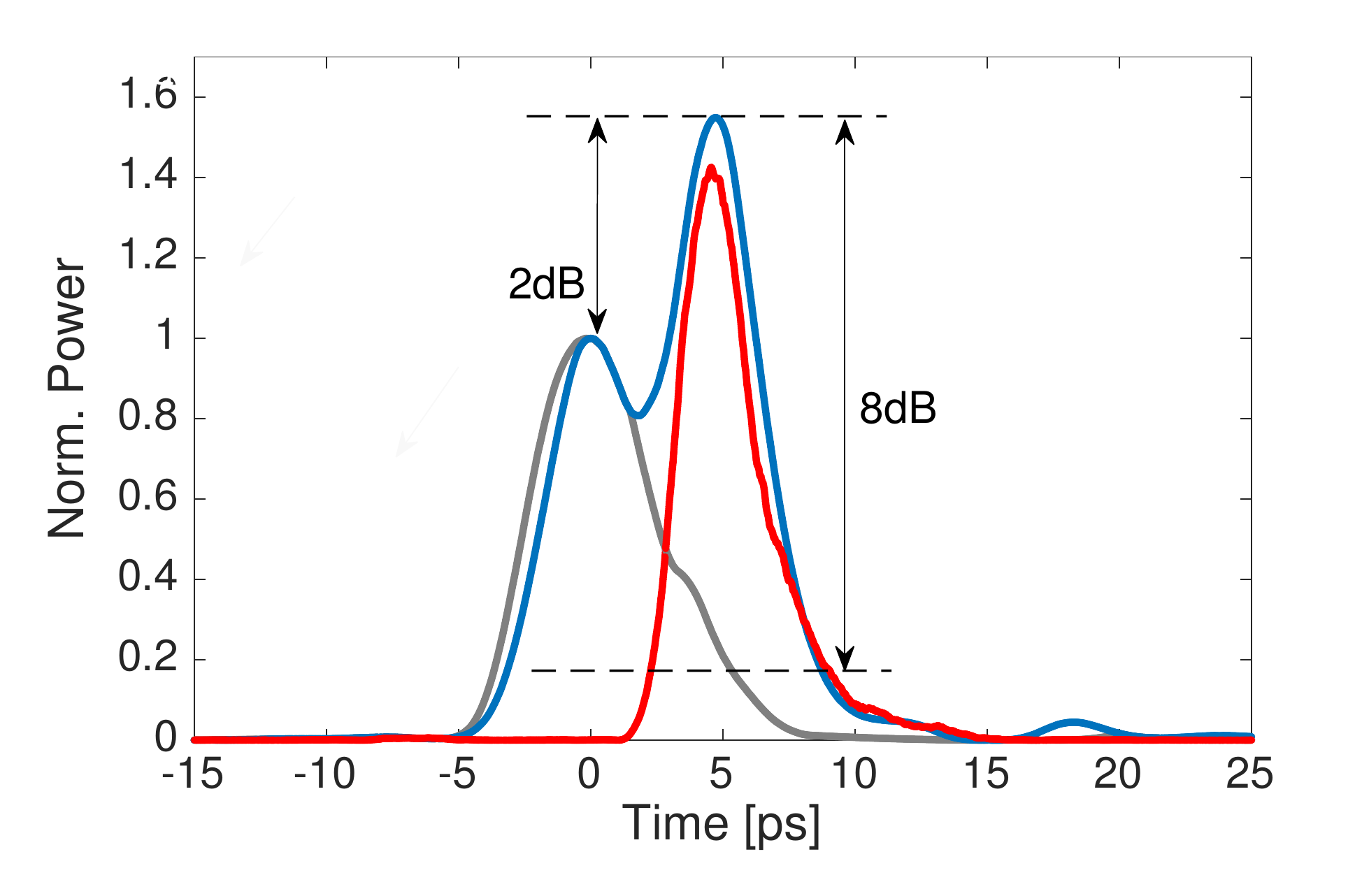}}
\caption{Measured temporal traces of the idler (red), as well as the probe with (blue) and without (grey) the presence of the pump. The pump power is about 120fJ/pulse and the pump-probe delay is set to -3~ps.}
\label{Fig_VI} 
\end{figure}

In order to estimate the gain level, we performed additional measurements on another sample where we recorded both the idler and the probe trace. The results are presented in \fig{Fig_VI}. 
Data presented in this figure would correspond to a vertical cut (Y-Axis) performed on both \fig{Fig_III}-a) and \fig{Fig_IV}-a), and provides a better view of the exact temporal dynamics of the probe/idler. The secondary peak is seen to coincide with the trace of the idler. In addition, the total gain is here as high as 8~dB and the amplified probe even exceeds by 2~dB the maximal reference probe amplitude at $t=0\unit{ps}$ when the pump is turned off. Moreover, the total probe energy has been increased by about 3~dB, which could not be obtained by pulse reshaping, further confirming that the secondary peak is a parametric-gain effect.  

Following these experimental observations and numerical analysis, we conclude that the parametric gain is directly related to i) the strength of the probe field in the cavity when the pump enters, and ii) the intrinsic efficiency of the gain mechanism (i.e. phasematching). As a result, the idler quickly vanishes as the pump-probe delay is decreased (delay $\le -4\unit{ps}$).

\begin{figure}
\centering
\includegraphics[width=8.5cm]{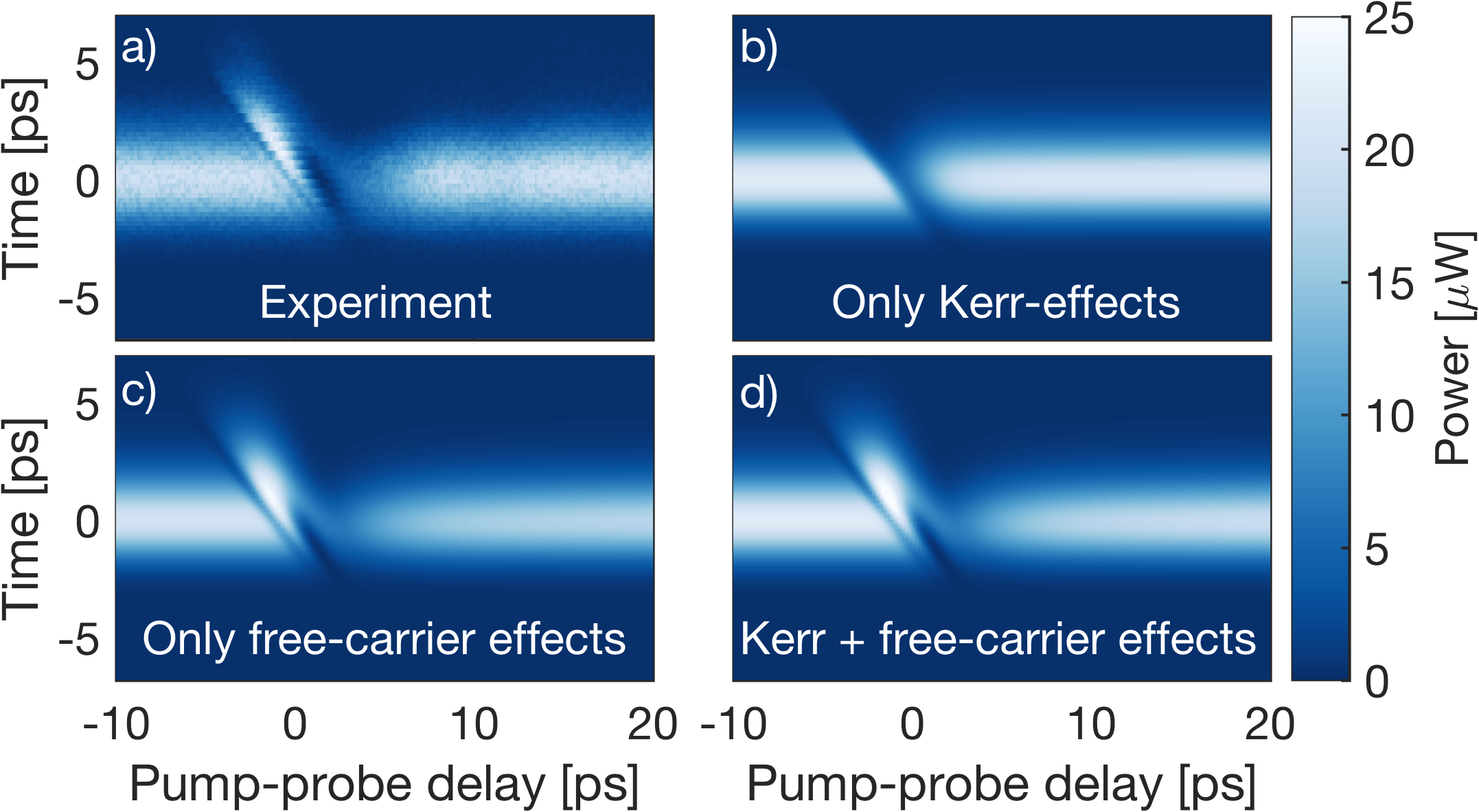}
\caption{a) Colormap of the probe temporal trace for various pump-probe delays. The pump power is about $130\unit{fJ/pulse}$. a) Experimental map b) Simulation wherein the FCD due to the fast free carriers plasma is neglected c) Simulation wherein the Kerr effect is turned off. d) full model simulation including both FCD and Kerr effects.}
\label{Fig_VII} 
\end{figure}

Finally, we investigated the origin of the parametric gain by comparing in \fig{Fig_VII} experimental data with numerical models taking into account coherent pump-probe mixing. It is clear that FCD is responsible for the parametric gain, while the Kerr effect has little influence on the overall dynamics. 
Thus, the weak probe and the strong pump pulse interference leads to a temporal modulation of the free-carrier generation rate. In turn, the coherent oscillation of the free-carrier population modulates the refractive index of the cavity via FCD, that scatters the strong pump into probe and idler. The bandwidth of the parametric gain (i.e. the admissible pump-probe detuning) is directly related to the lifetime of the free-carrier population, since the corresponding bandwidth must sustain the pump-probe beat-frequency. Thus, both fast diffusion and slower carrier recombination \cite{Yu:13} may be important, depending on the pump-probe-cavity detuning. A shorter free-carrier relaxation time thus supports parametric gain at higher detunings (beat-frequencies). 



In conclusion we have presented temporally resolved pump-probe transmission measurements of an ultrafast PhC all-optical switch. The use of a heterodyne cross-correlation detection scheme in combination with pulse shaping enabled the characterization of the cavity dynamics with unprecedented temporal and energy resolution. 
This resulted in the experimental observation of strong coherent dynamics, an effect previously neglected in the analysis of optical switches. In particular, four-wave mixing induced by oscillations of the spatially localized free-carriers in the nanocavity leads to more than 10~dB parametric gain of the probe and the generation of a strong idler signal. Such wave-mixing, induced by population oscillations, is 
already known in semiconductor optical amplifiers that are several hundred micrometers long \cite{uskov94,Mork97} but here the interaction takes place in a cavity with an extent that is a fraction of a wavelength and relies on cavity-enhanced two-photon absorption rather than normal linear absorption. Further understanding the role of such coherent dynamics is expected to be important for the development of photonic switches, possibly offering new regimes of ultra-fast switching based on coherent effects.



\section{Acknowledgments}
This work was supported by the Villum Foundation through the VKR center of excellence NATEC and the Danish Research Council for Independent Research (Grant No. FTP 11-116740). The authors thank M. Heuck for fruitful discussions and Kresten Yvind for the assistance on device fabrication.

\textbf{Contributions}: P.C and P.L. contributed equally to this work.


\bibliography{Bibliography}

\end{document}